# The Impact of Supervision and Incentive Process in Explaining Wage Profile and Variance

Nitsa Kasir (Kaliner)[a] and Idit Sohlberg[b],*

[a] *Research Department, Bank of Israel, Israel*
[b] *Department of Economics and Business Management, Ariel University, Israel*

The implementation of a supervision and incentive process for identical workers may lead to wage variance that stems from employer and employee optimization. The harder it is to assess the nature of the labor output, the more important such a process becomes, and the influence of such a process on wage development growth. The dynamic model presented in this paper shows that an employer will choose to pay a worker a starting wage that is less than what he deserves, resulting in a wage profile that fits the "classic" profile in the human-capital literature. The wage profile and wage variance rise at times of technological advancements, which leads to increased turnover as older workers are replaced by younger workers due to a rise in the relative marginal cost of the former.

Keywords: Supervision, Monitoring, Incentives, Wage developments, Technological changes.
JEL codes: C61, J31, M52, O33

We would like to thank Prof. Tsiddon from Tel-Aviv University for his guidance and helpful comments, our colleagues from the Bank of Israel and Ariel University for their useful advice and Gideon Israel for his editorial contribution.

*Corresponding author. Ariel University, Ariel 40700, Israel. Tel.: +972-524308199.
E-mail address: aditso@ariel.ac.il

# 1. Introduction

In widely used labor-market models, homogeneous workers who work under identical conditions receive identical wages. This outcome prevalent even when one shifts from the basic models of a competitive market and human capital to more complex models of bargaining and efficiency wage. According to these models, workers' wages are determined on the basis of their demographic indicators,[1] the characteristics of their enterprise or sector,[2] and institutional factors.[3] It has been found, however, that in the past two decades these indicators manage to explain only part of the wage-setting process.[4] In this paper, we use the theoretical framework of "supervision" and "efficiency wage" to explain the influence of supervision and incentive process on wage disparities, especially in time of technological advances.

Worker incentivization and supervision processes are fundamental to a firm's existence.[5] The relevant literature on this topic focuses mainly on the outcomes of this supervision, particularly its positive effect on the worker's level of efficiency and effort, and also on the adverse implications of supervision due to the crowding-out phenomenon.[6] The importance of personal supervision as opposed to team supervision,[7] and the desired supervision and hierarchical structure[8] are also important sub-topics in this literature. Given the development of production processes that yield hard-to-estimate outputs, the importance of incentivation and supervision has grown in recent decades, especially in sectors such as high-tech and financial services, where much is invested in employee evaluation and its outcome remuneration.

---

[1] Katz & Murphy, 1991; Juhn, Murphy, & Pierce, 1993; Acemoglu, 1996, 1998, 1999, 2002; Caselli, 1999; Galor & Moav, 2000; Aghion, Howitt, & Violante, 2002; David, Katz & Kearney. 2006.
[2] Davis & Haltiwanger , 1991; DiNardo, Fortin, & Lemieux, 1996; Entorf, & Kramarz, 1997; Lee, 1999; Neumark, 2000 .
[3] Card & DiNardo, 2002; Lemieux, 2008.
[4] Juhn, Murphy, & Pierce, 1993; Gosling, Machin, & Meghir, 2000; Violante, 2002; Baker & Solon, 2003; Autor, Katz & Kearney ,2008; Firpo, Fortin, & Limieux, 2011; Alvaredo et al., 2013.
[5] Mirrlees, 1974, Barzel, 1982; Eisenhardt,1989; Fama, 1991; Chandler, 1992; Fehr & Goette, 2007.
[6] Holmström & Milgrom, 1994; Prendergast, 1999; Strobl & Walsh, 2007; Dickinson & Villeval, 2008; Laffont & Martimort, 2009.
[7] Alchian & Demsetz, 1972; Holmström, 1982; .Fehr & Gächter, 2000; Van Dijk, Sonnemans, & Van Winden, 2001; Sefton, Shupp & Walker, 2007.
[8] Williamson, 1976; Calvo & Stanislaw 1979; Waldman,1984; Fumas, 1993; Qian, 1994; Meagher, 2001; Rajan & Wulf, 2006;. DeVaro & Waldman, 2012; Accard, 2014.

This paper emphasizes the link among the diverse parameters that define the supervision process, wage development and its variance. Workers in the model choose the level of effort that is optimal for them in light of the supervision level, starting wage, and level of incentivization, all of which are determined by a profit-maximizing employer. By solving the simultaneous model and analyzing the results over time, particularly at a time of technological change, we obtain an explanation for some of the unexplained increase in wage variance in recent decades.

In Section 2 of the paper, the framework of the model is presented and the worker's optimal behavior is examined. In Section 3, specific functions for wage and utility equations are defined. In Section 4, the model is completed with the inclusion of rules of optimal employer behavior and examination of how wage variance is explained by exogenous factors, such as cost of checking and technological improvements. Section 5 concludes.

**2. The Worker**

In the model described, the worker chooses his optimal level of effort at work. His employer may monitor this level by supervising his, but only at a cost. The utility of the representative worker in each period rises with an increase in consumption and decreases with an increase in effort. To simplify, let us assume that a worker has a additive[9] utility function that does not change during the period (much like the function presented by Calvo and Wellisz, 1978, 1979):

2.1 $\quad U_t = u(c_t) - v(e_t),$
$\quad\quad\quad u'(c_t) \geq 0, \quad v'(e_t) \geq 0, \quad u''(c_t) \leq 0, \quad v''(e_t) \geq 0,$
$\quad\quad\quad 0 \leq e_t \leq 1, \quad c_t \geq 0,$

---

[9] Below we also examine a Cobb-Douglas utility function, which we solve by means of dynamic numerical programming.

Where $c_t$ equals the consumption of a worker in period $t$ and $e_t$ represents the level of effort in period $t$, ($e_t = 1$ represents maximum effort and $e_t = 0$ represents total inactivity). We assume that in each period the individual consumes his entire wage and has no other sources of income.

Each worker has a potential work life of T periods. In each period, the worker maximizes his expected utility from that period onward:[10]

$$2.2 \qquad U^j = \sum_{t=j}^{T} \delta^{t-j} \cdot E(U_t) \qquad j = 1,....,T$$

where $\delta$ is the time discount factor, $0 < \delta < 1$.

Before hiring a worker, the employer discloses his payment structure and invites the employee to take it or leave it. The employer commits to pay the worker base wage $w_0$, which may increase if the worker demonstrates strong effort on the job or attains specific goals agreed upon with the employer. We assume that the employer will give similar workers identical base wages and that, since employees have the same preferences, they will choose the same initial level of effort.

In accordance with the employment contract, the employee has $p$ likelihood of being evaluated in each period, irrespective of whether or not he was evaluated in the past. If the worker is not evaluated during a certain period, his wage will remain at the previous period's level; if he is evaluated, it will be recalculated on the basis of his effort and the production that it yields. His bonus or penalty will be calculated commensurate with the gap between the wage he deserves and the wage he received in the last period. The wage agreement concluded with the worker may be defined for each period $t$ as follows:

---

[10] We assume that the worker makes no ab initio commitment to a future level of effort; instead, in each individual period he chooses the effort level that is optimum for his at this time.

$$2.3 \quad w_t = \begin{cases} w_{t-1} & \text{if not evaluated} \\ \max\{\tilde{w}(e_t) + \alpha \cdot (\tilde{w}(e_t) - w_{t-1}), 0\} & \text{if evaluated} \end{cases}$$

where:

$\tilde{w}'(e_t) \geq 0, \tilde{w}''(e_t) \leq 0$ and $0 \leq \alpha \leq 1$.

Due to the element of bonus in the wage function, this model belongs to the class of "efficiency wage" models, in which wage is adjusted (up or down) on the basis of effort expended. In this paper, it is assumed that, as opposed to other compensation structures that doom workers to dismissal if their productivity fails to meet predetermined expectations, our worker will keep his job but his wages will be cut.

Importantly, the employer can use only one employee as a test case with which to determine his expectations of other workers. If so, one may question the utility of sampling more than one worker. It may be assumed, however, that legal restrictions limit the employer's ability to revise the employee's wage unless he is personally sampled. The need for this type of employment contract has been researched and is usually associated with a state of insufficient information about the production process.[12][11,]

For presentation purposes, let us assume that there are two periods[13]. As previously mentioned, the employee's likelihood of being evaluated is represented by *p*. The employee's wage is dependent on whether he was evaluated in the current period on the basis of his previous wage, taking into consideration whether he was rewarded or penalized for his level of production. The dynamics of this process are shown in Figure 2.1.

---

[11] See, for example, Hashimoto, 1981, and Hall & Lazear, 1984. Alternately, one may assume the existence of several groups of workers and that, even though all workers sampled in each of the groups are homogeneous, the employer acts under conditions of uncertainty as to the type of worker sampled.

[12] It can be assumed that if the worker is evaluated his wage will be changed in accordance with his previous earnings. We found that the main results of the model, given the foregoing wage function, do not change. In an alternative wage structure, bonuses are construed as nonrecurrent wage increases. If a worker excels in a certain period, he receives a one-time bonus that does not affect his wage in the future. This type of structure, prevalent in the business world, is discussed below with the help of numerical dynamic programming.

[13] A comparative statistics analysis of the influence of the parameters on effort when there is only one period appears in Appendix 1.

**Figure 2.1: Development of Effort and Wage over Course of Two Periods**

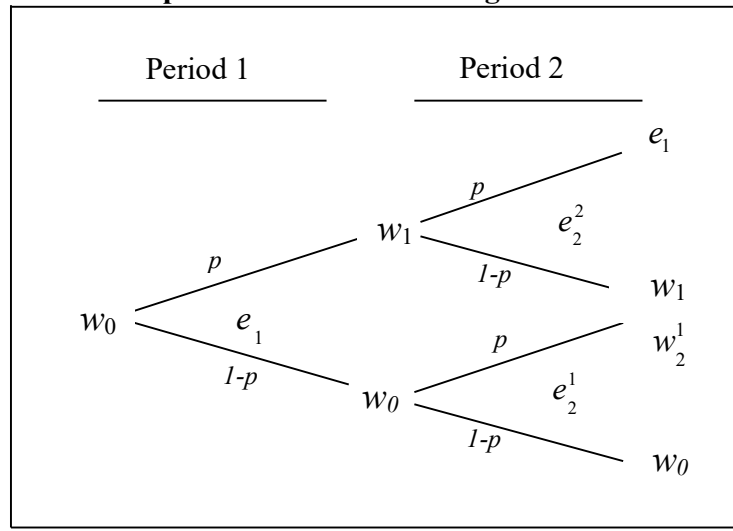

It may be seen that at the end of Period 2 the workers' wages spread in four different directions, reflecting the difference in their effort/production during this period and the sampling in both periods. In Table 2.1, the various wage levels are shown in accordance with the process defined in Figure 2.1:

**Table 2.1: Distribution of Wages after Two Periods**
(In accordance with the definitions in Figure 1.1)

| Wage | Calculation | Probability |
| --- | --- | --- |
| $w_0$ | $w_0$ | $(1-p)^2$ |
| $w_1$ | $\widetilde{w}(e_1) + \alpha \cdot (\widetilde{w}(e_1) - w_0)$ | $p \cdot (1-p)$ |
| $w_2^1$ | $\widetilde{w}(e_2^1) + \alpha \cdot (\widetilde{w}(e_2^1) - w_0)$ | $p \cdot (1-p)$ |
| $w_2^2$ | $\widetilde{w}(e_2^2) + \alpha \cdot [\widetilde{w}(e_2^2) - (\widetilde{w}(e_1) + \alpha \cdot (\widetilde{w}(e_1) - w_0))]$ | $p^2$ |

We now examine the employee's considerations recursively. We assume that he has reached the beginning of Period 2, that he knows the wage that he received in Period 1, and that he uses this knowledge to choose his level of effort in Period 2. In other words, he will maximize his effective expectancy in Period 2:

2.4 $\quad U^2 = p \cdot u(w_2) + (1-p) \cdot u(w_{t-1}) - v(e_2).$

If the employee was sampled in Period 1, $w_2 = w_2^2$, $w_{t-1} = w_1$, $e_2 = e_2^2$; if not, then: $w_2 = w_2^1$, $w_{t-1} = w_0$, $e_2 = e_2^1$.

Let us assume that the effort level in Period 2 is the optimal level of effort in this period:

2.5 $\quad e_2^* = \arg\max_{e_2} U^2.$

The first-order condition for the maximization of the utility function yields optimal rules of behavior of the employee:

2.6 $\quad p = \dfrac{v'(e_2^*)}{(1+\alpha) \cdot \widetilde{w}'(e_2^*) \cdot u'(w_1)}$

At the beginning of Period 1, the employee knows the terms of his optimal behavior in Period 2 and will maximize his expected utility function for both periods:

2.7 $\quad \begin{aligned} U^1 = {} & [p \cdot u(w_1) + (1-p) \cdot u(w_0) - v(e_1)] + \\ & + \delta \cdot p \cdot [p \cdot u(w_2^2) + (1-p) \cdot u(w_1) - v(e_2^2)] + \\ & + \delta \cdot (1-p) \cdot [p \cdot u(w_2^1) + (1-p) \cdot u(w_0) - v(e_2^1)]. \end{aligned}$

The optimal amount of effort for the employee in Period 1 will defined as:

2.8 $\quad e_1^* = \arg\max_{e_1} U^1$

and, arising from the first-order condition for maximizing the utility function, we get the employee's optimal rules of behavior:

$$2.9 \quad p = \frac{v'(e_1^*)}{(1+\alpha)\cdot \widetilde{w}'(e_1^*)\cdot \left[ u'(w_1) + \delta\cdot((1-p)\cdot u'(w_1) - p\cdot \alpha \cdot u'(w_2^2))\right]}$$

At the optimum, the ratio of the marginal utility of leisure to the marginal utility of consumption is fixed and equal to the sampling rate. For a deeper analysis of the relations among the variables, a specific definition of the utility and wage functions is needed. This is provided in the next section.

## 3. Characterization of Utility and Wage Functions

### 3.1 Additive Utility Function in Consumption and Effort

Let us assume that the utility of consumption is logarithmic and that the utility of effort is proportionate to this level, i.e.:

$$3.1 \quad U_t = \ln(w(e_t)) - b\cdot e_t$$

where $b$ is the parameter.

To simplify this, let us define function $\widetilde{w}(e_t)$ as equal to the employee's effort:[14]

$$3.2 \quad \widetilde{w}(e_t) = e_t$$

And consequently the wage function is presented for each period $t$ as:

$$3.3 \quad w_t = \begin{cases} w_{t-1} & \text{if not evaluated} \\ \max\{e_t + \alpha\cdot(e_t - w_{t-1}), 0\} & \text{if evaluated} \end{cases}$$

where $0 \leq \alpha \leq 1$.

Within this framework, if the employee maintains a constant level of effort over time and the employer evaluates his progress in each work period, his bonuses or penalties will be partly offset in future periods. If, however, his effort levels gradually increase from period to period, his wage will also increase in each future period (Figure 3.1).

---

[14] Calvo and Wellisz (1979) assumed, for the function defined in Equation 3.2, that the effort level is multiplied by a constant. At this stage, for simplicity's sake, we assume that this constant is equal to 1. This assumption will be removed for reasons spelled out in Part 4 below.

**Figure 3.1:**
**Wage Development of Worker Who Is Sampled in Each Period**[a]

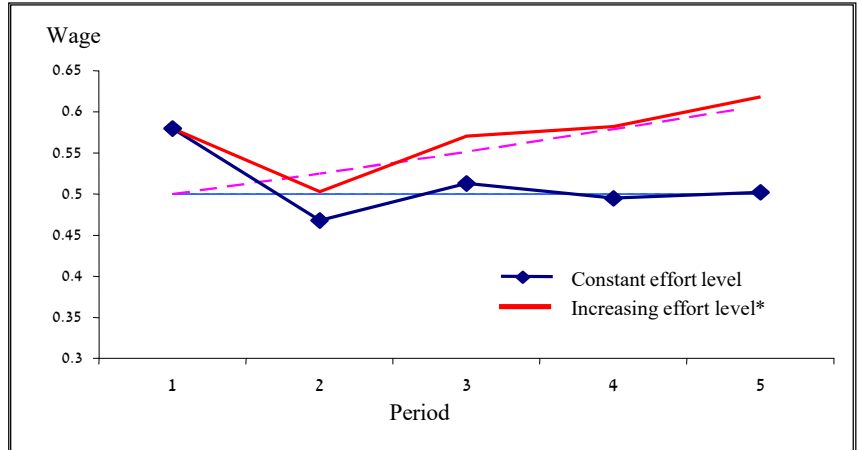

[a] Assuming: $w_0=0.5$, $e_0=0.3$, $\alpha=0.4$, $p=1$, $T=5$. We assume that the effort level is exogenous, an assumption that will be removed in the next paragraph.
* Effort increases by 5% each period.

The expected utility of a worker in Period 1 is:

$$3.4 \quad U^1 = \sum_{t=1}^{T} \delta^{t-1} \cdot$$

A Bellman equation is used to find the optimal effort level - $e_t^*$:

$$3.5 \quad g(e_{T-t}) = \max_{e_{T-t}} \left\{ \left[ p \cdot \ln(e_{T-t} + \alpha \cdot (e_{T-t} - w_{T-t-1})) + (1-p) \cdot \ln(w_{T-t-1}) \right] - b \cdot e_{T-t} \right] + \delta \cdot g(e_{T-t-1}) \right\}.$$

and the optimal level of effort in each period $t$ is:

$$3.6 \quad e_t^* = \frac{p}{b} \cdot \varphi_t + \frac{\alpha}{1+\alpha} \cdot w_{t-1},$$

where: 
$$\varphi_t = \frac{\sum_{t}^{T}(\delta \cdot (1-p))^{T-t}}{1 + \alpha \cdot \delta \cdot p \cdot \sum_{t}^{T}(\delta \cdot (1-p))^{T-t-1} - \frac{\alpha \cdot p}{1-p}}$$

Coefficient $\varphi$ decreases over time because, as time passes, fewer and fewer work periods remain in which a level of effort can influence wage decreases (when $t = T$, we get $\varphi$ in its minimal value, equal to 1). Thus, the level of effort also decreases over time. By plugging the optimal effort level (Formula 3.6) into the wage formula (Formula 3.3), we get the wage level for each period $t$:

$$3.7 \quad w_t = \begin{cases} w_{t-1} & \text{if not evaluated} \\ \frac{p}{b} \cdot (1+\alpha) \cdot \varphi_t & \text{if evaluated} \end{cases}$$

where $w_{t-1}$ in Period 1 is the base wage ($w_0$).

We find that the wage that the employee will receive if sampled is independent of his past wage and dependent only on the parameters of the models.[15,16] This independence significantly narrows the wage possibilities that exist at the end of each period, making it relatively easy to calculate wage distribution even for multiple periods at a time. For example, the worker will receive in period $T$ the wage that is determined for Work Period 2 – $p \cdot (1+\alpha) \cdot \varphi_2$ – at probability $p \cdot (1-p)^{T-2}$. After $T$ periods, there will be ($T+1$) wage possibilities, distributed as follows:

---

[15] This is due to the definition of the utility function, in which the marginal utility of idleness is constant and the wage function is proportional to effort.

[16] For simplicity's sake, the model presented shows a Marcov process, in which the wage paid to a worker in each period is a dependency of the previous period only, irrespective of the earlier past. Alternatively, a more complex structure may be presented, in which wage in each period is a dependency of the worker's employment history, e.g., the number of periods that passed since his latest sampling. Such models provide additional explanations for the wage scatter but do not affect the main results.

**Table 3.1: Wage Distribution**

| Wage | Last period in which worker was evaluated and wage set | Probability of wage in Period $T$ |
|---|---|---|
| $w_0$ | Was never evaluated | $(1-p)^T$ |
| $p \cdot (1+\alpha) \cdot \varphi_t$ | 1<br>2<br>.<br>t<br>.<br>T-1 | $p \cdot (1-p)^{T-t}$ |
| $p \cdot (1+\alpha)$ | T | $p$ |

Figure 3.2 below describes the development of wage over time as the outcome of the worker's level of effort. The figure shows the wage expectancy according to the distribution defined in Table 3.1 and in extreme cases where the worker is sampled in each period.

The wage expectancy profile described in the figure resembles the "classic wage profile", in which wage rises at an accelerated rate in the first few years and then begins to settle slowly or even to decrease. This wage profile and, specifically, the accelerated increase in wage in the first few years, stem from the base wage being too low for the worker's level of effort, $w_0 < \widetilde{w}(e_t)$. If the employer decides to compensate the worker so that $w_0 = \widetilde{w}(e_t)$, the wage profile would be completely horizontal; if the employer overcompensates the employee, $w_0 > \widetilde{w}(e_t)$, the wage profile will begin to decrease from the outset.

**Figure 3.2:**

**Development of Wage**[a]

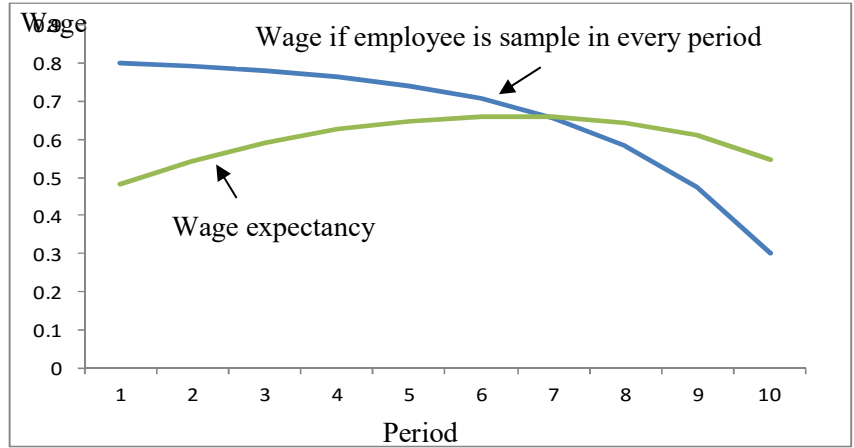

[a] Assuming: $w_0=0.4, \alpha=0.5, p=0.2, b=1, \delta=0.90, T=10$

To analyze the modularity of the effect on wage variance, we first check the results for an individual during one period. In accordance with Formula 3.6, the optimal effort level – $e*$ – that a worker will choose is:

$$3.8 \qquad e* = \frac{p}{b} + \frac{\alpha}{1+\alpha} \cdot w_o .$$

Hence:

$$3.9 \qquad e* = e(\overset{+}{p}, \overset{-}{b}, \overset{+}{\alpha}, \overset{+}{w_0})$$

For simplicity, it is assumed from now on that $b=1$.

That is, an increase in the sampling rate, the size of the bonus (penalty), and the starting wage will induce an increase in the worker's level of effort during the work period. Conversely, an increase in the utility coefficient for idleness will induce a decrease in the worker's level of effort.

The wage function for one period is

$$3.10 \qquad w = \begin{cases} w_o & \text{if not evaluated} \\ p \cdot (1+\alpha) & \text{if evaluated} \end{cases}$$

and the wage variance at the end of the period is:

3.11 $\quad Var(w) = p \cdot (1-p) \cdot [p \cdot (1+\alpha) - w_O]^2$.

In other words, the worker's level of effort, wage profile, and wage variance are all defined by the parameters included in the contract offered to his (Equation 2.8). Take, for example, the wage variance described in Figure 3.3 as a function of two parameters in the employment contract: base wage and bonus/penalty rate. It may be easily seen that wage variance will be larger when one's base wage is lower than one is worth $w_0 > \tilde{w}(e)$, and smaller when base wage exceeds worth $w_0 < \tilde{w}(e)$. When the behavior of wage variance is examined as a function of other parameters in the model, similar outcomes are obtained. Now let us assume that the employee is employed for more than one work period. Much like the one-period model and even more so the multi-period model, wage variance depends on the relation between the parameters that determine the employment contract as decided upon by the employer.

**Figure 3.3:**
**Wage Variance for One Period as a Function of Probability of Evaluation and for Different Values of Beginning Wage $w_0$**

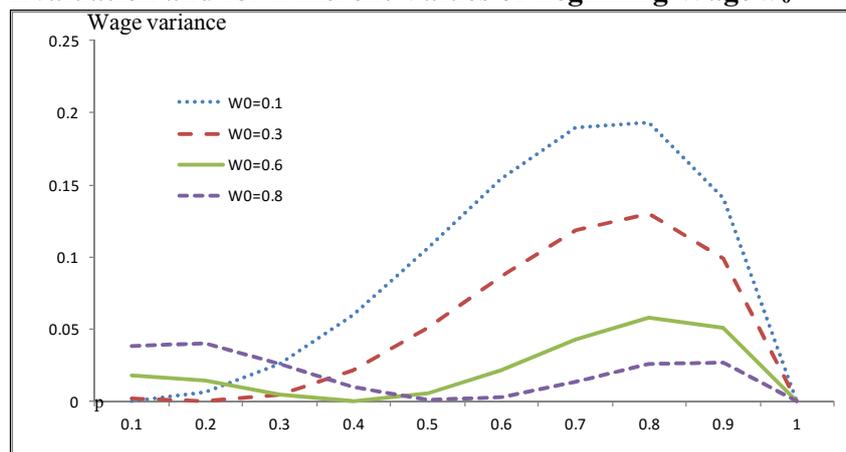

[a] Assuming: $b=1, \alpha=0.5, T=1$

## 3.2 Cobb-Douglas Production Function

In this section we assume that worker's utility function for period $t$ is Cobb-Douglas and not additive as above:[17]

3.12 $\quad U_t = (1-e_t)^\gamma \cdot c_t^\beta$

where the variables are as defined above and $\beta$ and $\gamma$ are parameters:

We also assume here that in a period when the worker is evaluated, he receive a nonrecurrent bonus that is not considered part of wage and will not affect his wage if he is evaluated again.[18] This means that in each period $t$ the worker receives wage $w_t$ and bonus $B_t$, defined as follows:

3.13 $\quad w_t = \begin{cases} w_{t-1} & \text{if not evaluated} \\ e_t & \text{if evaluated} \end{cases}$

3.14 $\quad B_t = \begin{cases} 0 & \text{if not evaluated} \\ \alpha \cdot (e_t - w_{t-1}) & \text{if evaluated} \end{cases}$

and consumption in each period is paid by the worker's wage and bonus in that period:

3.15 $\quad c_t = w_t + B_t$

Due to the difficulty in applying an analytical solution, we perform this analysis by means of numeric dynamic programming, in which the worker chooses the optimal effort level in each period, taking into consideration the future repercussions of his decisions. Assume that there are ten periods. The worker's optimal effort level in each period is defined in Table 3.2 (minimum = 0; maximum = 1). For example, a worker who was paid 0.4 in Period 5 will choose

---

[17] This function replaces the additive utility function set forth in Equation 3.1.
[18] This definition of the bonus would have no qualitative effect on the outcomes in the utility function set forth in the previous section.

an effort level of 0.5. We can see that the outcome of the Cobb-Douglas production function, much like that of the additive function, the optimal effort level decreases in tandem with the expected number of periods. Unlike the earlier example, however, in the Cobb-Douglas production function a higher wage in a previous period will causes a decrease in future effort level

**Table 3.2:**
**Optimal Effort Level in Ten Work Periods as a Function of Wage in Previous Period[a]**

| Period: | 1 | 2 | 3 | 4 | 5 | 6 | 7 | 8 | 9 | 10 |
|---|---|---|---|---|---|---|---|---|---|---|
| 0   | 1   | 1   | 1   | 1   | 1   | 1   | 1   | 0.9 | 0.9 | 0.6 |
| 0.1 | 0.9 | 0.9 | 0.8 | 0.8 | 0.8 | 0.8 | 0.8 | 0.7 | 0.5 | 0   |
| 0.2 | 0.8 | 0.7 | 0.7 | 0.7 | 0.7 | 0.7 | 0.6 | 0.5 | 0.3 | **0** |
| 0.3 | 0.7 | 0.6 | 0.6 | 0.6 | 0.6 | 0.6 | 0.5 | 0.4 | **0.2** | 0 |
| 0.4 | **0.6** | 0.6 | **0.6** | 0.5 | **0.5** | 0.5 | **0.4** | **0.3** | 0.1 | 0 |
| 0.5 | 0.5 | 0.5 | 0.5 | 0.5 | 0.4 | **0.4** | 0.3 | 0.2 | 0.1 | 0 |
| 0.6 | 0.4 | **0.4** | 0.4 | **0.4** | 0.4 | 0.3 | 0.3 | 0.2 | 0.1 | 0 |
| 0.7 | 0.3 | 0.3 | 0.3 | 0.3 | 0.3 | 0.3 | 0.2 | 0.2 | 0.1 | 0 |
| 0.8 | 0.3 | 0.3 | 0.3 | 0.3 | 0.2 | 0.2 | 0.2 | 0.1 | 0.1 | 0 |
| 0.9 | 0.2 | 0.2 | 0.2 | 0.2 | 0.2 | 0.2 | 0.1 | 0.1 | 0.1 | 0 |
| 1   | 0.1 | 0.2 | 0.2 | 0.2 | 0.2 | 0.1 | 0.1 | 0.1 | 0.1 | 0 |

(Wage in previous period is the row label.)

[a] Assuming: $\alpha=0.5$, $T=10$, $p=0.2$, $\alpha=0.1$, $\gamma=0.3$, $\beta=0.7$, $\delta=0.95$

A worker's wage development over time is an outcome of his effort level. Take, for example, the extreme case in which the worker is sampled in each period and given starting wage ($w_0$) 0.4. The effort track that the worker chooses is highlighted (the number is underlined) in Table 3.2. The wage and bonus that he will receive as a result are described in Table 3.3. Looking at the table, it is easy to see that effort and wage development are not monotonous in the first six work periods, whereas from the seventh period both begin to decrease over time due to the finite time horizon.

**Table 3.3:**
**Optimal Effort Level, Wage and Bonus Level for Ten Work Periods**[a]

| Period: | 1 | 2 | 3 | 4 | 5 | 6 | 7 | 8 | 9 | 10 |
|---|---|---|---|---|---|---|---|---|---|---|
| Optimal effort | 0.6 | 0.4 | 0.6 | 0.4 | 0.5 | 0.4 | 0.4 | 0.3 | 0.2 | 0 |
| Wage (if sampled) | 0.6 | 0.4 | 0.6 | 0.4 | 0.5 | 0.4 | 0.4 | 0.3 | 0.2 | 0 |
| Bonus (if sampled) | 0.02 | 0.02 | -0.02 | 0.02 | 0.01 | -0.01 | 0 | -0.01 | -0.01 | - |
| Total compensation | 0.62 | 0.42 | 0.58 | 0.42 | 0.51 | 0.39 | 0.4 | 0.29 | 0.19 | 0 |

[a] Assuming: $p = 0.2, \alpha = 0.1, \gamma = 0.3, \beta = 0.7, \delta = 0.95, w_0 = 0.4$

In the previous example, we chose an extreme case in which the worker is sampled in each work period. In reality, a worker is sampled, and in turn his wage distribution in future periods is determined, with a probability of *p*. An example of wage distribution in each of ten periods (where *p=0.2*) is presented in Table 3.4. The probability presented in each cell in the table is for all wage values in a defined category.

**Table 3.4:**
**Wage Distribution for Ten Work Periods**[a]

| Period: | 1 | 2 | 3 | 4 | 5 | 6 | 7 | 8 | 9 | 10 |
|---|---|---|---|---|---|---|---|---|---|---|
| **Wage***: | | | | | | | | | | |
| 0-0.1 | | | | | | | | | | 0.15 |
| 0.1-0.2 | | | | | | | | | 0.03 | 0.07 |
| 0.2-0.3 | | | | | | | | 0.05 | 0.09 | 0.22 | 0.18 |
| 0.3-0.4 | | | | 0.04 | 0.10 | 0.16 | 0.19 | 0.20 | 0.30 | 0.26 | 0.21 |
| 0.4-0.5 | 1.00 | 0.80 | 0.64 | 0.51 | 0.51 | 0.54 | 0.54 | 0.44 | 0.35 | 0.28 |
| 0.5-0.6 | | 0.20 | 0.32 | 0.39 | 0.33 | 0.27 | 0.21 | 0.17 | 0.14 | 0.11 |
| **Total probability:** | 1.00 | 1.00 | 1.00 | 1.00 | 1.00 | 1.00 | 1.00 | 1.00 | 1.00 | 1.00 |

[a] Assuming: $p = 0.2, \alpha = 0.1, \gamma = 0.4, \beta = 0.6, w_0 = 0.4, \delta = 0.95$

\* The table shows the total probability of receiving a wage within the defined wage bracket from the lowest value (not inclusive) to the highest (inclusive).

The development of wage expectancy and variance are presented in Figure 3.4, which describes the variables for a timeframe of ten and twenty periods. Here as in the previous section, wage expectancy behaves in a manner similar to the development of the classic wage profile, ascending in the first periods with a decreasing slope and decreasing later. An increase in wage variation over time was also found.

**Figure 3.4:**
**Wage Expectancy and Variance for Cobb-Douglas Production Function**[a]

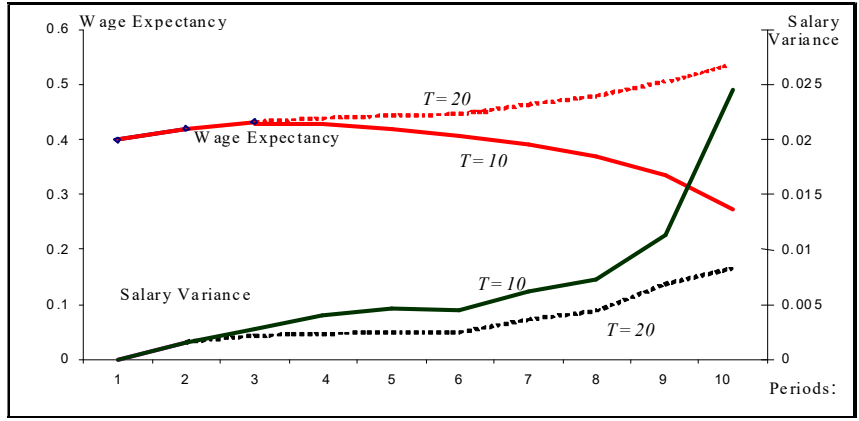

[a] Assuming: $\delta = 0.95$, $p = 0.2$, $\alpha = 0.1$, $\gamma = 0.4$, $\beta = 0.6$, $T = 10, 20$

## 4. The Employer

We complete the model by adding the optimal maximization of the employer. The output of the worker in period $t$ is commensurate with his level of effort and is not dependent on other workers' output and effort (an additive assumption). The production function ($y_t$) is defined as follows:

4.1 $\quad y_t = f(e_t),$
$\quad\quad\quad f'(e_t) \geq 0, \quad f''(e_t) \leq 0.$

The employer benefits from the revenues that his employees' labor generates in each period. His revenue function (scaled to one worker) is:

4.2 $\quad \pi = \sum_{t=1}^{T} \eta^{t-1} \cdot \left[ f'(e_t) - (E(w_t) + p \cdot c) \right]$

where $\eta$ is the employer's discount factor.

Using first-order conditions, differentiation with respect to $p$ yields:

4.3 $\quad \sum_{t=1}^{T} \eta^{t-1} \cdot (f'(e_t) \cdot \frac{de_t}{dp} - \frac{dE(w_t)}{dp} - c) = 0$

and differentiation with respect to $w_o$ yields:

$$4.4 \qquad \sum_{t=1}^{T} \eta^{t-1} \cdot (f'(e_t) \cdot \frac{de_t}{dw_0} - \frac{dE(w_t)}{dw_0}) = 0$$

and differentiation with respect to $\alpha$ yields:

$$4.5 \qquad \sum_{t=1}^{T} \eta^{t-1} \cdot (f'(e_t) \cdot \frac{de_t}{d\alpha} - \frac{dE(w_t)}{d\alpha}) = 0$$

Defining p*, $w_0$*, and α* as optimal parameters for employers who satisfy equations 4.3–4.5, we examine their influence on the worker's effort level, wage, and wage variance. For simplicity's sake, we first examine the model for one period. We assume that the producer produces under a constant return to scale. This function is suitable for a fixed wage per unit of effort, much like the wage function defined in Section 3:

4.6 $\qquad f(e) = k \cdot e$,

And the discounted revenue for the employer is:

$$4.7 \qquad \pi(e_t) = \sum_{t=1}^{T} \eta^{t-1} \cdot \left\{ k \cdot e_t - \left[ p \cdot (\tilde{w}(e_t) + \alpha \cdot (\tilde{w}(e_t) - w_{t-1})) + (1-p) \cdot w_0 + p \cdot c \right] \right\}$$

We insert into the revenue function (Formula 4.7) the wage function defined in Section 3 and the optimal effort level that the worker will choose for the additive productivity function:

$$4.8 \qquad \tilde{w}(e_t) = e_t \qquad e_t = p + \frac{\alpha}{1+\alpha} \cdot w_{t-1}.$$

What we get is that the employer will act – optimally – according to the following rules:

$$4.9 \qquad \alpha^* = \frac{\sqrt{k \cdot c}}{k-1} - 1$$

$$4.10 \qquad p^* = 1 - \frac{\alpha^*}{1+\alpha^*} \cdot k = \frac{(1-k) \cdot (\sqrt{c} - \sqrt{k})}{\sqrt{c}}$$

4.11 $\quad w_0^* = \dfrac{[(1+a^*)\cdot p^*]^2}{k} = (\sqrt{k}-\sqrt{c})^2$

We find that at the optimum, the employer will choose a lower base wage than what the worker deserves for his efforts ($w_0^* < (1+\alpha)\cdot p = \widetilde{w}(e)$). This policy will lead to different wages among identical workers and wage expectations similar to the "classical" wage profile (as seen in Figure 3.2). An increase in supervision cost will induce a decrease in the optimum employee sampling rate ($p$) to a raise in optimum base wage ($w_0$) and relative bonus ($\alpha$). These will lead to decreases in the worker's effort level, wage, and wage variance.

Examining the effect of an increase in marginal production (technological advancements) on wage expectancy and variance is more complicated because in this case the worker is influenced by two factors – an increase in effort level due to a change in the parameters of his labor contract and an increase in exchange per unit of effort due to a marginal increase in output for that unit. We assume that wage as a function of effort will be determined by the following (this definition of wage function is an expansion of the function defined in Equation 3.3)[19]:

4.12 $\quad \widetilde{w}(e) = (\lambda \cdot k)\cdot e$,

and $\lambda<1$, a choice which integrates models of specific human capital where the revenue is divided between the employer and employee.[20]

An increase in marginal production will lead to an increase in the employer's optimum sampling rate and optimum base wage paid. The rate of the optimal bonus/penalty, however, will decrease. Due to the large gap between the wage that the employee deserves for his efforts and his base wage the total bonus/penalty amount will increase despite its rate decrease.

---

[19] See note 16 above.
[20] In a general equilibrium solution, the optimal $\lambda$ may be tested for. This solution, however, exceeds the purpose of this paper.

Simultaneously, an increase in marginal output will induce increases in the employee's level of effort,[21] his wage expectancy, and wage variance among employees (Figure 4.1).

**Figure 4.1: Wage Expectancy and Variance as a Function of Marginal Production for One Period**[a]

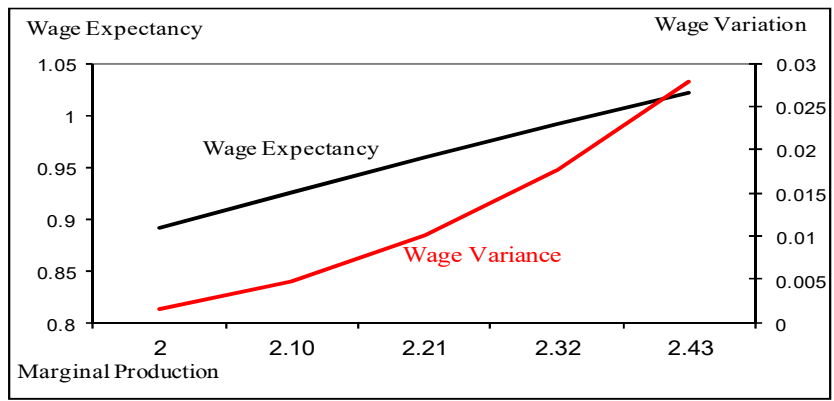

[a] *Assuming: C=0.2, $\lambda$ =0.8*

An analysis of the effect of technological advancements on optimal employee and employer policies over time is complicated and necessitates the use of numerical planning. This plan calculates the sampling amount, the base wage, and the relative size of the bonus that maximizes the value of discounted revenues. By using these optimal parameters, we can find the worker's optimal long-term effort level, wage expectancy, and wage variance.

We find that technological advancements raise the wage profile over time due to (a) an increase in wage per unit of effort, (b) an increase in the sampling rate, and (c) an increase in the effort level, in both the additive utility and the Cobb-Douglas utility functions. Technological changes will also cause wage variance to rise due to the rise in sampling rate and the gap in salary between sampled workers and un-sampled worker.

We also found that technological advancement will, raise the cost of employing older workers (calculated by subtracting their marginal output from wage expectancy) relative to newer workers[22] (Figure 4.2). This result may provide an additional explanation for the increase in the turnover rate of young workers replacing older workers, even if their other characteristics are the same.

---

[21] Furthermore, the normalized SD coefficient in wage expectancy will rise.

**Figure 4.2:**
**Development of Wage Expectancy and Variance before and after Technological Advancements**[a]
Utility function $U_t = \ln(w(e_t)) - b \cdot e_t$

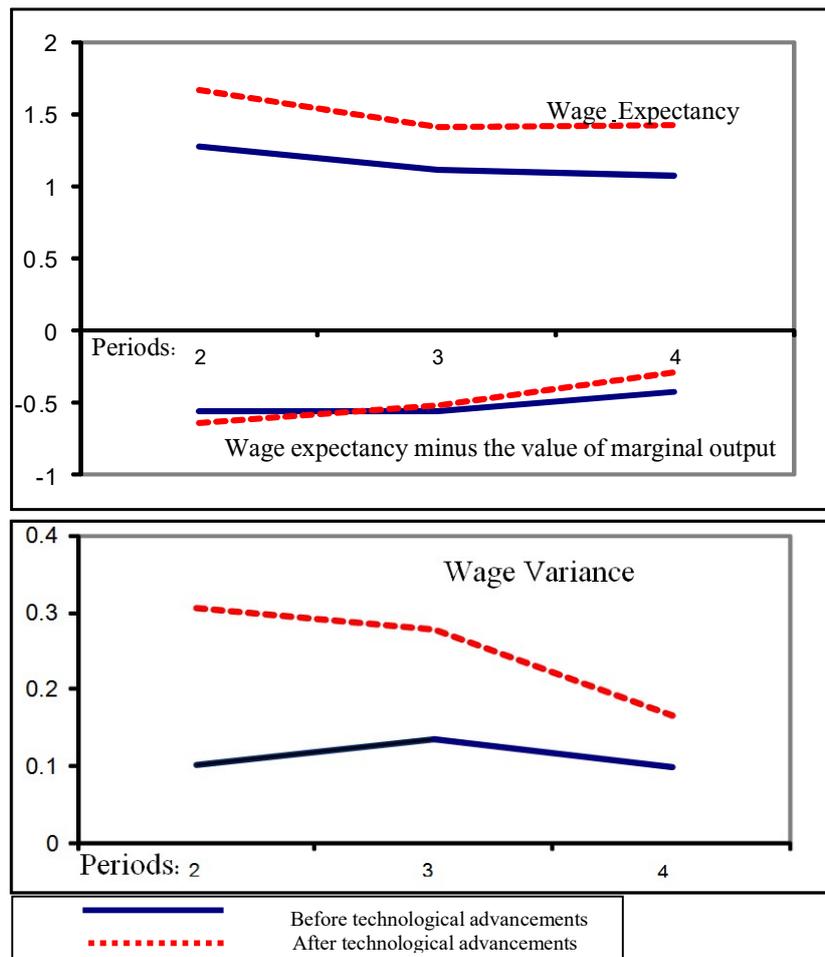

[a] *Assuming:* $C=0.2$, $\lambda=0.8$. For simplicity of presentation we show periods 2 to 4.

---

[22] This condition obtains when the cost of training new workers is low enough. In this paper, we assume zero training cost.

**5. Conclusion**

In the model described, an employer's evaluation system, comprising of incentives and penalties, causes different wage development among identical workers. The incentive and punitive measures are proportional to the worker's effort level in relation to his wage. As a result of the optimization of workers' and employers' behaviors, wage differences among workers develop inevitably because the employer will choose to pay his workers a starting wage that is less than their starting effort level.

As a result of the dynamic structure of the model, workers will choose to lower their level of effort in their later periods in the labor market. For this reason and due to the sampling process, their wage expectancy profile will fit the "classic" profile described in the human-capital literature – rising quickly in the first work periods, gradually evening out, and falling in later periods.

At times of technological advancements, wage expectancy and variance will grow. Additionally, it will also increase the gap between the worker's wage and his marginal output value in a way that motivates the employer to replace older workers with younger ones.

In this paper we offer an alternative explanation for some of the most important characters of the labor market in recent decades. In times of technological advancements, when it is difficult to estimate output and production, explanations that focus on the supervision and incentive process increase in importance.

# References

Accard, Philippe. (2014). Complex hierarchy: The strategic advantages of a trade-off between hierarchical supervision and self-organizing. European Management Journal.

Acemoglu, D. (1996). A microfoundation for social increasing returns in human capital accumulation. Quarterly Journal of Economics, 111(3), 779-804.

Acemoglu, D. (1998). Why do new technologies complement skills? Directed technical change and wage inequality. Quarterly Journal of Economics, 113(4), 1055-1090.

Acemoglu, D. (1999). Changes in unemployment and wage inequality: An alternative theory and some evidence. American Economic Review, 89(5), 1259-1278.

Acemoglu, D. (2002). Technical change, inequality, and the labor market. Journal of Economic Literature, 40(1), 7-72.

Alchian, A., & Demsetz, H. (1972). Production, information costs, and economic organization. The American Economic Review, 62, 777-795.

Alvaredo, F., Atkinson, A., Piketty, T., & Saez, E. (2013). The top 1 percent in international and historical perspective. Journal of Economic Perspective, 27, 3-20.

Aghion, P., Howitt, P., & Violante, G.L. (2002). General purpose technology and wage inequality, Journal of Economic Growth, 7(4), 315-45.

Autor, D.H., Katz, L.F., & Kearney, M.S. (2008). Trends in US wage inequality: Revising the revisionists. The Review of Economics and Statistics, 90.2, 300-323.

Baker, M., & Solon, G. (2003). Earnings dynamics and inequality among Canadian men, 1976-1992: Evidence from longitudinal income tax records. Journal of Labor Economics, 21(2), 289-321.

Barzel, Y. (1982). Measurement cost and the organization of markets. Journal of Law and Economics, 27-48.

Calvo, G.A., & Stanislaw, W. (1978). Supervision, loss of control and the optimum size of the firm. Journal of Political Economy, 86(5).

Calvo, G.A., & Stanislaw, W. (1979). Hierarchy, ability and income distribution. Journal of Political Economy, 87(5), 991-1010.

Caselli, F. (1999). Technological revolutions. American Economic Review, 89, 78-102.

Card, D., & DiNardo, J.E., (2002). Skill-biased technological change and rising wage inequality: Some problems and puzzles. Journal of Labor Economics, 20/4, 733-783.

Chandler, A.D. (1992). Organizational capabilities and the economic history of the industrial enterprise. Journal of Economic Perspectives, 79-100.

Davis, S. J., & Haltiwanger, J. (1991). Wage dispersion between and within US manufacturing plants, 1963-1986. National Bureau of Economic Research (No. w3722).

David, H., Katz, L. F., & Kearney, M. S. (2006). The polarization of the US labor market. National Bureau of Economic Research (No. w11986).

DeVaro, J., & Waldman, M. (2012). The signaling role of promotions: Further theory and empirical evidence. Journal of Labor Economics, 30(1), 91-147

Dickinson, D., & Villeval, M.C. (2008). Does monitoring decrease work effort? The complementarity between agency and crowding-out theories. Games and Economic Behavior, 63, 56-76.

DiNardo, J., Fortin, N., & Lemieux, T. (1996). Labor market institutions and the distribution of wages, 1973-1992: A semi-parametric approach. Econometrica, 64, 1001-1044.

Eisenhardt, K. M. (1989). Agency theory: An assessment and review. Academy of management review, 14(1), 57-74.

Entorf, H., & Kramarz, F. (1997). Does unmeasured ability explain the higher wages of new technology workers?. European Economic Review, 41(8), 1489-1509.

Fama, E. (1991). Time, salary and incentive payoffs in labor contracts. Journal of Labor Economics, 9, 25-44.

Fehr, E., & Gächter, S. (2000). Cooperation and punishment in public goods experiments. American Economic Review, 90(4), 980-994.

Fehr, E., & Goette, L. (2007). Do workers work more if wages are high? Evidence from a randomized field experiment. American Economic Review, 97(1), 298-317.

Firpo, S., Fortin, N. M., & Lemieux, T. (2011). Occupational tasks and changes in the wage structure. Discussion paper series//Forschungsinstitut zur Zukunft der Arbeit. (No. 5542)

Fumas, V. S. (1993). Incentives and supervision in hierarchies. Journal of Economic Behavior & Organization, 21(3), 315-331.

Galor, O., & Moav, O. (2000). Ability-biased technological transition, wage inequality within and across groups, and economic growth. Quarterly Journal of Economics, 115(2), 469-497.

Gosling, A., Machin, S., & Meghir, C. (2000). The changing distribution of male wages in the UK. Review of Economic Studies, 67(4), 635-666.

Hall, R.E., & Lazear, E.P. (1984). The excess sensitivity of layoffs and quits to demand. Journal of Labor Economics, 2(2), 233-257.

Hashimoto, M. (1981). Firm-specific human capital as a shared investment. The American Economic Review, 475-482.

Holmström, B. (1982). "Moral hazard in teams. Bell Journal of Economics, 13, 324-340.

Holmström, B., & Milgrom, P. (1994). The firm as an incentive system. The American Economic Review, 84, 972–991.

Juhn, C., Murphy, K.M., & Pierce, B. (1993). Wage inequality and the rise of returns to skill. Journal of Political Economy, 101, 410-442.

Katz, L. F., & Murphy, K. M. (1991). Changes in relative wages, 1963-1987: Supply and demand factors (No. w3927). National Bureau of Economic Research.

Laffont, J.J., & Martimort, D. (2009). The theory of incentives: The principal-agent model. Princeton University Press.

Lee, D. (1999). Wage inequality in the U.S. during the 1980s: Rising dispersion or falling minimum wage? Quarterly Journal of Economics, 64, 977-1023.

Lemieux, T. (2008). The changing nature of wage inequality. Journal of Population Economics, 21/1, 21-48.

Meagher, K. J. (2001). The impact of hierarchies on wages. Journal of Economic Behavior & Organization, 45(4), 441-458.

Mirrlees, J. (1974). The optimal structure of incentives and authority within an organization. Bell Journal of Economics, 7(1), 105-131.

Neumark, D. (2000). Changes in job stability and job security: A collective effort to untangle, reconcile, and interpret the evidence. National bureau of economic research (No. w7472).

Prendergast, C. (1999). The provision of incentives in firms. Journal of Economic Literature, 37(1), 7-63.

Qian, Y. (1994). Incentives and loss of control in an optimal hierarchy. The Review of Economic Studies, 61(3), 527-544.

Rajan, R. G., & Wulf, J. (2006). The flattening firm: Evidence from panel data on the changing nature of corporate hierarchies. The Review of Economics and Statistics, 88(4), 759-773.

Sefton, M., Shupp, R., & Walker, J. (2007). The effect of rewards and sanctions in provision of public goods. Economic Inquiry, 45, 671-690.

Strobl, E,. & Walsh, F. (2007). "Estimating the shirking model with variable effort." Labour Economics 14.3, 623-637.

Van Dijk, F., Sonnemans, J., & van Winden, F. (2001). Incentive systems in a real effort experiment. European Economic Review, 45, 187-214.

Violante, G.L. (2002). Technological acceleration, skill transferability and the rise in residual inequality. Quarterly Journal of Economics, 117(1), (2002), 297-338.

Waldman, M. (1984). Worker allocation, hierarchies and the wage distribution.The Review of Economic Studies, 51(1), 95 -109

Williamson, Oliver, E. (1967). Hierarchial control and optimum firm size. Journal of Political Economy, 75(2), (1967), 123-138.

**Appendix 1:**
**Single-Period Model: The influence of the supervision parameters on the effort level.**

The utility function is:

(1) $U = p \cdot u(\widetilde{w}(e) + \alpha \cdot (\widetilde{w}(e) - w_0)) + (1-p) \cdot u(w_0) - v(e)$

and the first-order conditions yield:

(2) $p = \dfrac{v'(e)}{u'(\widetilde{w}(e) + \alpha \cdot (\widetilde{w}(e) - w_0)) \cdot (1+\alpha) \cdot \widetilde{w}'(e)}$

Define:

(3) $\psi(e) = (1+\alpha) \cdot \widetilde{w}(e)$,

We get:

(4) $p = h(e, \alpha, w_0) \equiv \dfrac{v'(e)}{u'(\psi(e) - w_0) \cdot \psi'(e)}$

we derive according to *e* and get:

(5) $h'_e = \dfrac{u' \cdot \psi' \cdot v'' - v' \cdot [u'' \cdot \psi'^2 + u' \cdot \psi'']}{u'(\psi(e) - w_o) \cdot \psi'(e))^2} > 0$

Under the assumptions concerning the derivations defined in Equation 2.1, we find that an increase in e induces an increase in *h(e)*.

<u>Comparative statistics:</u>

1) Assuming that $p \uparrow \Rightarrow h(e) \uparrow$ (from Equation 4), it follows that $\uparrow e$ (from Equation 5).

2) Assuming that $w_0 \uparrow \Rightarrow h(e) \downarrow$ (from Equation 4), it follows that(from Equation 5). $\uparrow e$

3) Assuming that $\alpha \uparrow$, it may be seen from Equation 3 that if $\widetilde{w}(e) \leq w \Rightarrow h \downarrow \Rightarrow e \uparrow$ - with rise in the rate of penalty, the level of effort rises, and if $\widetilde{w}(e) > w_0 \Rightarrow ? h(?) \Rightarrow e(?)$ - with a rise in the rate bonus, the direction of the effect is not clear.